\documentclass[aps,prl,reprint,groupedaddress]{revtex4-1}

\usepackage{graphicx}
\usepackage{dcolumn}
\usepackage{bm}
\usepackage{amsbsy}
\usepackage{graphicx}
\usepackage{hyperref}
\usepackage{graphics}
\usepackage{mathrsfs}

\newcommand{\be}{\begin{equation}} \newcommand{\ee}{\end{equation}}
\newcommand{\bea}{\begin{eqnarray}} \newcommand{\eea}{\end{eqnarray}}

\newcommand{\re}[1]{(\ref{#1})}

\newcommand{\fig}[1]{figure \ref{#1}}
\newcommand{\brt}[1]{[#1]}
\newcommand{\para}{\paragraph}

\renewcommand{\a}{\alpha}
\renewcommand{\b}{\beta}

\renewcommand{\d}{\delta}

\newcommand{\rmd}{\mathrm{d}}

\newcommand{\nonum}{;\ }
\newcommand{\etal} {et al.}
\newcommand{\ie}{i.e.\ }

\newcommand{\adot}{\dot{a}}

\newcommand{\OKn}{\Omega_{K0}}

\renewcommand{\l}{\mathrm{l}}
\newcommand{\s}{\mathrm{s}}
\newcommand{\dls}{d_\mathrm{ls}}
\newcommand{\dl}{d_\mathrm{l}}
\newcommand{\ds}{d_\mathrm{s}}
\newcommand{\tE}{\theta_{\mathrm{E}}}

\newcommand{\PRD}[1]{{\it Phys. Rev.} {\bf D#1}}

\newcommand{\PRL}[1]{{\it Phys. Rev. Lett.} {\bf #1}}

\newcommand{\PLB}[1]{{\it Phys. Lett.} {\bf B#1}}
\newcommand{\MNRAS}[1]{{\it Mon. Not. Roy. Astron. Soc.} {\bf #1}}
\newcommand{\APJ}[1]{{\it Astrophys. J.} {\bf #1}}

\newcommand{\CQG}[1]{{\it Class. Quant. Grav.} {\bf #1}}
\newcommand{\GRG}[1]{{\it Gen. Rel. Grav.} {\bf #1}}
\newcommand{\AaA}[1]{{\it Astron. \& Astrophys.} {\bf #1}}

\begin{document}

\title{A new test of the FLRW metric using the distance sum rule}

\author{Syksy R\"{a}s\"{a}nen}
\affiliation{University of Helsinki, Department of Physics, and Helsinki Institute of Physics, P.O. Box 64, FIN-00014 University of Helsinki, Finland}

\author{Krzysztof Bolejko}
\affiliation{Sydney Institute for Astronomy, School of Physics, A28, The University of Sydney, Sydney, NSW 2006, Australia}

\author{Alexis Finoguenov}
\affiliation{University of Helsinki, Department of Physics, P.O. Box 64, FIN-00014 University of Helsinki, Finland}

\date{\today}

\begin{abstract}

We present a new test of the validity of the
Friedmann--Lema\^{\i}tre--Robertson--Walker (FLRW)
metric, based on comparing the distance from redshift
0 to $z_1$ and from $z_1$ to $z_2$ to the distance from $0$ to $z_2$.
If the universe is described by the FLRW metric, the comparison
provides a model-independent measurement of spatial curvature.
The test relies on geometrical optics, it
is independent of the matter content of the universe and the
applicability of the Einstein equation on cosmological scales.
We apply the test to observations, using the Union2.1 compilation
of supernova distances and Sloan Lens ACS Survey
galaxy strong lensing data.
The FLRW metric is consistent with the data, and the spatial
curvature parameter is constrained to be $-1.22<\OKn<0.63$,
or $-0.08<\OKn<0.97$ with a prior from the cosmic microwave
background and the local Hubble constant, though modelling of
the lenses is a source of significant systematic uncertainty.

\end{abstract}


  
\maketitle

\setcounter{tocdepth}{2}

\setcounter{secnumdepth}{3}

\section{Introduction} \label{sec:intro}

\para{Testing the FLRW metric.}

In addition to providing tight constraints on cosmological
parameters in specific models, the increasing precision
and breadth of cosmological observations makes it possible
to test assumptions behind entire classes of models.
A particularly important assumption is that the universe 
is, on average, described by the exactly homogeneous and isotropic
Friedmann--Lema\^{\i}tre--Robertson--Walker (FLRW) metric.
More precisely, we consider the assumption that light propagation
over long distances is described by the FLRW metric.
This can be tested by consistency conditions between different
observables derived from geometrical optics. Such tests are independent of
the matter content of the universe and its relation to spacetime geometry
(usually given by the Einstein equation).
It has been proposed that the observed late-time accelerated expansion
could be related to the failure of the FLRW approximation.
Possibilities include extra dimensions \cite{Ferrer},
violation of statistical homogeneity and isotropy \cite{LTB},
and the effect of deviation from exact homogeneity and isotropy on
the average expansion rate, \ie
backreaction \cite{BR, Boehm:2013}.

Testing the FLRW metric by comparing observations
of the expansion rate and distance was proposed in \cite{Clarkson:2007b}
and implemented in \cite{Shafieloo:2009, Mortsell:2011, Sapone:2014}.
A similar test using parallax distance and angular diameter distance
was put forth in \cite{Rasanen:2013}.
We propose a third consistency test, based on the sum rule of
distances along null geodesics of the FLRW metric, and apply it to real data.
If the sum rule is violated, the FLRW metric is ruled out.
If the data is consistent with the sum rule, the test provides a
model-independent measurement of the spatial curvature of the universe,
like the tests proposed in \cite{Clarkson:2007b, Rasanen:2013}.

\section{FLRW consistency condition} \label{sec:cons}

\para{Distances.}

If space is exactly homogeneous and isotropic, spacetime
is described by the FLRW metric
\bea \label{metric}
  \rmd s^2 = - \rmd t^2 + \frac{a(t)^2}{1 - K r^2} \rmd r^2 + a(t)^2 r^2 \rmd\Omega^2 \ ,
\eea

\noindent where $K$ is a constant related to the spatial curvature;
the Ricci scalar of the hypersurface of constant proper time is
$6 K/a(t)^2$.
(When $K>0$, the metric \re{metric} covers only half of the spacetime.)
The Hubble parameter is $H\equiv\adot/a$, and
its present value is denoted by $H_0$.
Let $D_A(z_\l,z_\s)$ be the angular diameter distance of a source at
redshift $z_\s$ (emission time $t_\s$) as seen at redshift $z_\l$
(observation time $t_\l>t_\s$). From \re{metric}, we find that the
dimensionless distance $d(z_\l,z_\s)\equiv (1+z_\s) H_0 D_A(z_\l,z_\s)$ is
\bea \label{bidt}
  d(z_\l, z_\s) &=& \frac{1}{\sqrt{-k}} \sinh\left( \sqrt{-k}\int_{t_\s(z_\s)}^{t_\l(z_\l)} \frac{H_0 \rmd t}{a(t)} \right) \ ,
\eea

\noindent where $k\equiv K/H_0^2$. We denote $d(z)\equiv d(0,z)$.

\para{Distance sum rule.}

Using \re{bidt}, $\dls\equiv d(z_\l,z_\s)$  can be written
in terms of $\dl\equiv d(z_\l)$ and $\ds\equiv d(z_\s)$ as
\bea \label{sum}
  \dls &=& \epsilon_1 \ds \sqrt{1 - k \dl^2} - \epsilon_2 \dl \sqrt{1 - k \ds^2} \ ,
\eea

\noindent where $\epsilon_i=\pm1$. For $k\leq0$, $\epsilon_i=1$.
For $k>0$, the signs depend on which halves of the three-dimensional
hypersphere the source and the lens are located and in which
direction the light propagates.
If there is a one-to-one correspondence between $t$ and $z$
and $d'(z)>0$, then $\epsilon_i=1$. We assume that this is the case,
so we have
\bea \label{sumrat}
  \frac{\dls}{\ds} &=& \sqrt{1 - k \dl^2} - \frac{\dl}{\ds} \sqrt{1 - k \ds^2} \ .
\eea

\noindent The relation \re{sum} (or \re{sumrat}) is a sum rule
for distances in the FLRW universe. 
(The case \re{sumrat} is given in e.g. \cite{Peebles:1993}, p. 336.)
In the spatially flat case, the distances are simply added together,
whereas for non-zero spatial curvature the relation is more involved.
Using \re{sumrat} in the case $|k|\ll1$ to obtain a model-independent
estimate of the spatial curvature was proposed in \cite{Bernstein:2005}.

\para{The consistency condition.}

The sum rule \re{sum} has been derived from the FLRW metric.
We get a consistency condition by solving for $k$ to obtain
(for all $\epsilon_i$)
\bea \label{k}
  k_S &=& - \frac{\dl^4 + \ds^4 + \dls^4 - 2 \dl^2 \ds^2 - 2 \dl^2 \dls^2 - 2 \ds^2 \dls^2}{4 \dl^2 \ds^2 \dls^2} \ ,
\eea

\noindent where the subscript $S$ indicates
that $k$ has been solved from the sum rule for distances.
We now drop the assumption that the universe is described
by the FLRW metric and take \re{k} as the definition of
a function $k_S(z_\l, z_\s)$ in any spacetime
(neglecting angular dependence).
If the universe is described by the FLRW metric, $k$ is constant
and equal to $-\OKn$, the present value of the spatial
curvature density parameter.
If it is observationally found that $k_S$
is different for any two pairs $(z_\l,z_\s)$, then light
propagation on large scales is not described by the FLRW metric.
The converse is not true: if $k_S$ is constant, this
does not imply that the metric is FLRW.

The consistency condition \re{k} provides a powerful test. In principle,
the FLRW metric can be falsified by measuring the three quantities
$(\dl, \ds, \dls)$ for two different values of $(z_\l, z_\s)$.
The test is very general, because it assumes only geometrical optics
and that light propagation can be described with the FLRW metric.
Unlike the condition between distance and expansion
\cite{Clarkson:2007b}, the consistency condition \re{k}
does not involve derivatives of the distance.
Unlike the condition between angular diameter and parallax distances
\cite{Rasanen:2013}, there are already measurements of the distances
involved, $d$ and $\dls$, on cosmological scales.

\section{Determining $d$ and $\dls$} \label{sec:det}

\para{The distance $d$.}

The Union2.1 compilation \cite{Suzuki:2011}
provides luminosity distances $D_L$ to 580 supernovae (SNe),
with arbitrary overall normalisation.
The highest redshift in the compilation is 1.4.
Normalising by $H_0$, we obtain $d_L\equiv H_0 D_L=(1+z) d$,
where the last relation holds in any spacetime \cite{duality}.
Our test involves only ratios of distances, so it does not
depend on the value of $H_0$.

In the Union2.1 analysis, the parameters that describe
SN light curves are fitted at the same time as the cosmological
parameters, and it is assumed that the universe is described by the
spatially flat FLRW model with dust and vacuum energy,
so the resulting distances are model-dependent
\cite{Nadathur:2010}. There are also significant differences between
light curve fitters \cite{fitter}. However, such effects
are likely subdominant to the uncertainties in the modelling
of the strong lensing systems that we use to determine $\dls$.
We therefore simply use the distances to SNe given in \cite{Suzuki:2011},
with the reported statistical and systematic errors.

\para{The distance $\dls$.}

Angular separation between strongly lensed images of the same source
depends on $\dls/\ds$ and the structure of the lens.
We assume that general relativity holds on the scale of the lensing system. 
If the lens can be approximated as a singular isothermal
ellipsoid (SIE), we have \cite{Kochanek:2004}
\bea \label{SIE}
  \frac{\dls}{\ds} &=& \frac{\tE}{4\pi f^2 \sigma^2} \ ,
\eea

\noindent where $\tE$ is the Einstein radius (in radians),
$\sigma$ is the velocity dispersion of the lens and
$f$ is a phenomenological coefficient that parametrises uncertainty due
to difference between the velocity dispersion of the observed stars and the
underlying dark matter, and other systematic effects.
Observations suggest the range $0.8<f^2<1.2$ \cite{Kochanek:1999, Ofek:2003}.

We consider two different treatments of \re{SIE},
which we call models Ia and Ib.
In model Ia, we take $f=1$.
In model Ib, we model $f$ by assigning an extra
Gaussian error of 20\% on $\dls/\ds$.
Leaving $f$ as a free parameter would significantly degrade the constraints
owing to a degeneracy between $f$ and $k$ for $-k\gg1$, due to
limited redshift coverage and small number of lensing data points.

We also consider a more complicated treatment of the lens,
introduced in \cite{Schwab:2009}, where \re{SIE} is replaced by
$\frac{\dls}{\ds} = N(\a,\b,\d) \frac{\tE^{\a-1}}{4\pi\sigma^2}$,
with $\a, \b$ and $\d$ being the slope of the density,
anisotropy of the velocity dispersion and the luminosity, respectively.
We call this model II.
Following \cite{Schwab:2009}, we treat $\a$ and $\b$ as
universal parameters with a Gaussian distribution with fixed
mean and variance. For $\d$ we use values reported for each
individual lens. These depend on the aperture.
We treat this variation as a lens-specific error on $\delta$,
assumed to be Gaussian, with the 1$\sigma$
range given by the difference between the maximum
and minimum values. The average mean value is
$\d=2.39$ and the average 1$\sigma$ error is 0.05.
The values given in \cite{Schwab:2009} are not centered around the
SIE model, due to non-zero mean anisotropy in the velocity
dispersion and the different slopes of the density and luminosity.
For the mean values of $\a$ and $\b$ and the value $\d=2.4$ used
in \cite{Schwab:2009}, $\dls/\ds$ is 12\% lower than in the SIE
case \re{SIE}. In our best-fits, the mean value of
$\dls/\ds$ in model II is 8\% lower than in model Ia,
and 9\% lower than in model Ib.

\para{Lensing data.}

We select strong lensing systems for which there is a
well-measured value for $z_\l$, $z_\s$, $\tE$ and $\sigma$.
We require the lens to be either
an elliptical or a lenticular galaxy,
so that it can be modelled as a SIE. We also require that there is either
an Einstein ring or arcs, not just multiple images, because without
individual spectra, we cannot be sure that separated images are from
the same source.
These criteria leave us with 30 lenses, listed in table II \cite{table2}.
We have checked that the lenses are isolated from other
galaxies and clusters.
The data is mostly from the Sloan Lens ACS Survey \cite{Bolton:2008},
with additional data from the SIMBAD database \cite{SIMBAD}
and \cite{Schwab:2014}.
The maximum source redshift is $z_\s=0.98$, well below the
maximum redshift of 1.4 of the Union2.1 SN compilation.
Following \cite{Bolton:2008}, we assign an error of 2\% on $\tE$
and a minimum error of 5\% on $\sigma$.

\section{Datafit and results} \label{sec:res}

\para{Fitting function.}

In principle, the function $k_S(z_\l, z_\s)$ defined in \re{k}
can be reconstructed from observations, and if it is not constant,
the FLRW metric is ruled out. Such a procedure has
been applied to $k$ defined with the expansion rate and distance
\cite{Clarkson:2007b} in \cite{Shafieloo:2009, Sapone:2014}.
However, \re{k} gives a biased estimate of $k$. If we insert
values of $\dl$, $\ds$ and $\dls$ with errors into \re{k},
it will not be centred on the real value of $k$.
In any case, given the small number of $\dls/\ds$ datapoints,
we do not try to find $k$ as a function of redshifts.
Instead, we fit a constant $k$ to the data and consider
the goodness of fit. Large values of $\chi^2$/d.o.f.
would be evidence against the FLRW model, or for unaccounted
errors.
If the FLRW hypothesis is not rejected, the $\chi^2$ values
give the probability distribution of $k$.

We obtain $d(z)$ and $k$ model-independently by
fitting to the SN and lensing data simultaneously.
As the fitting function for $d(z)$, we have compared polynomials of
different order, as well as splines, rational functions and
B\'ezier curves by fitting to mock datasets of FLRW
models with zero, positive or negative spatial curvature, as well as
the real data. We find that with current data, it doesn't make
much difference which function we use, as long as it is more
flexible than a second order polynomial.
Note that, in contrast to attempts to reconstruct
the deceleration parameter or the equation of state \cite{recon},
we do not need derivatives of $d$.
We present the results for a fourth order polynomial.
Because $d(0)=0, d'(0)=1$, it has three parameters.
Our fitting model thus consists of a fourth order polynomial
for $d(z)$, and $\dls/\ds$ given by \re{sumrat} with a constant $k$,
with four parameters in total.

\para{Upper limit on $k$ from CMB and $H_0$.}

On a hypersphere, the comoving angular diameter distance
is bounded from above by $1/\sqrt{K}$, so $k\leq1/d(z)^2$
for all $z$, and this applies our $k$ defined by the sum rule \re{sumrat}.
Given $d'>0$, the strongest constraint comes from the largest value of $z$.
We adopt the model-independent distance
$D_A(0,1090)=12.8\pm0.07$ Mpc from the cosmic
microwave background (CMB) \cite{CMB}
and the locally measured Hubble parameter
$H_0=72.5\pm2.5$ km/s/Mpc \cite{Efstathiou:2013}.
(We give error bars as 68\% limits and ranges as 95\% limits.)
These values do not depend on the assumption that the universe
at late times is well-described by the FLRW metric on large scales.
Taking the 2$\sigma$ lower bound for both quantities, we have $d>3.1$,
which implies $k<0.10$. (In fact, the conservative bounds $D_A>12$ Mpc and
$H_0>60$ km/s/Mpc would be sufficient for $k<0.1$.)

\begin{table}
\begin{center}
\begin{tabular}{|c|cccccc|c|}
\hline
Model &  Best-fit & Mean & 95\% range & $\chi_{\mathrm{SN}}^2$ & $\chi_{\mathrm{L}}^2$ & $\chi_{\mathrm{tot}}^2$ & Prior on $k$ \\
\hline
Ia  & 0.55 & 0.38 & [-0.63, 1.22] & 545 & 23 & 568 & None \\
Ib  & 0.76 & 0.37 & [-1.42, 1.54] & 545 & 20 & 565 & None \\
II  & -1.34 & -2.06 & [-5.98, 0.61] & 545 & 11 & 556 & None \\
\hline
Ia  & 0.09 & -0.25 & [-0.97, 0.08] & 545 & 28 & 573 & $k<0.1$ \\
Ib  & 0.09 & -0.53 & [-1.92, 0.07] & 545 & 21 & 566 & $k<0.1$ \\
II  & -1.34 & -2.29 & [-6.07, 0.13] & 545 & 11 & 556 & $k<0.1$ \\
\hline
\end{tabular}
\end{center}
\caption{Results for $k$ for different lens models,
with and without the prior from CMB and $H_0$.
In $\chi^2$, the subscript SN refers to SNe, L refers to lenses
and tot refers to both.
The number of SN datapoints is 580. The number of lensing datapoints
is 23 for model Ia and 30 for models Ib and II.}
\label{tab:k}
\end{table}

\para{Probability distribution for $k$.}

The $\chi^2$ for the SN data is the same for all three lens models,
but for the lensing data the $\chi^2$  is 78, 20 and 11 for models
Ia, Ib and II, respectively.
Given that we have 30 lensing datapoints,
model Ia underestimates the statistical errors, there are systematic
problems with the lensing data or the FLRW metric does not apply.
In any case, the increased errors of model Ib and the greater complexity
of model II seem to overcompensate.
A look at outliers indicates that the issue is probably systematics.
For model Ia there are 7 lenses that are outliers at more than 2$\sigma$.
If the reason was problems with the FLRW metric, we would expect
the outliers to show a distinct pattern in redshifts.
Instead, they are distributed randomly, as we would expect
if the problem is unmodelled systematic issues with the lenses.
A 2D Kolmogorov-Smirnov test indicates that for model Ia,
the probability that the outliers and non-outliers are
drawn from the same distribution is 29\% \cite{Press:2007}.
In fact, the lenses' goodness of fit does not form a pattern
for any of the lensing models.
We therefore conclude that the data does not provide evidence
for deviations from the FLRW metric.
We produce a conservative truncated list of lenses by removing the
most extreme outlier, refitting and iterating until all lenses are
within 2$\sigma$. This leaves us with 23 lenses for model Ia.
For models Ib and II we use all 30 lenses, as none are outliers.

\begin{figure}[t]
\scalebox{0.7}{\includegraphics[angle=0, clip=true, trim=0.0cm 0.0cm 0.0cm 0.0cm]{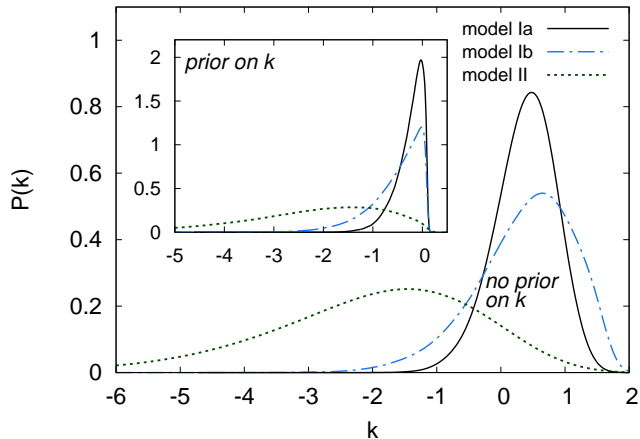}}
\caption{
Probability distribution of $k$, using the SIE model (black, solid),
SIE model with additional 20\% error (blue, dot-dashed) and the model of
\cite{Schwab:2009} (green, dotted). The inset shows the 
case with the prior $k<0.1$.
}
\label{fig:P}
\end{figure}

We marginalise over the three polynomial coefficients to obtain
the probability distribution $P(k)$.
The results are shown in \fig{fig:P}.
Even without a prior on $k$, the probability distribution is not
Gaussian, and has a tail at negative values of $k$.
The 95\% ranges, mean and best-fit values for $k$ as well as
the goodness-of-fit values are given in table \ref{tab:k}.
Our studies of mock datasets show that for current data the typical
offset of the mean and the best-fit from the real underlying values
is much smaller than the error bars.
Imposing the prior on $k$ leads to an increase in $\chi^2$
for model Ia, indicating tension between the lensing
and CMB data. There is no such issue in model II, but given its
large number of parameters, and possible systematic issues with
the lenses, it would be premature to conclude that it is more realistic.

For model Ia we have $-0.63<k<1.22$, or $-0.97<k<0.08$ with the prior.
For model Ib, the range increases by a factor of 1.6. For model II,
there is a long negative tail, due to degeneracy between
decreasing $N$ and increasing $-k>0$.
There is no evidence for spatial curvature.

For comparison, if we assume that the universe has FLRW geometry,
that the Einstein equation holds and that the matter consists
of dust and vacuum energy, we obtain 
$-0.53<k<0.52$ for models Ia and Ib and $-0.55<k<0.51$ for model II.
In this case, the constraints on $k=-\OKn$ are dominated by the SN data,
the lensing data is unimportant.

\section{Conclusions} \label{sec:conc}

\para{Results and comparison to previous work.}

The $k_S$ test based on the distance sum rule \re{sum}
for the FLRW metric is independent of the matter content
of the universe and its relation to spacetime geometry
on cosmological scales, though
general relativity has been assumed to be valid
when determining $\dls/\ds$ from astrophysical
systems.
We find that the data is consistent with the FLRW metric.
Treating lenses as SIE with the published errors and eliminating
outliers, the spatial curvature parameter $\OKn=-k$ is determined to be
$-1.22<\OKn<0.63$ from SN and lensing data, and 
$-0.08<\OKn<0.97$ when we include a prior from CMB and $H_0$.
These numbers are sensitive to lens modelling.

This range is two orders of magnitude wider than the one quoted
from the latest CMB plus baryon acoustic oscillation data,
$-0.007<\OKn<0.006$ \cite{Planck:cosmo}.
That assumes that the universe is described by a FLRW model
whose late-time matter content is dust and vacuum energy,
and that the Einstein equation is valid on cosmological scales.
However, tight constraints, $-0.007<\OKn<0.01$,
are also obtained in an analysis with loose priors on dark energy,
combining WMAP7 CMB, Union2 SN and Big Bang
Nucleosynthesis data as well as local  $H_0$ measurements
\cite{Okouma:2012}.
Without the $H_0$ value, which is debated
\cite{Efstathiou:2013, Busti:2014, Heavens:2014},
the constraint is  $-0.12<\OKn<0.01$.
The sensitivity is due to two distinct effects.
First, the overall angular scale of the CMB anisotropy pattern
provides a measurement of the angular diameter distance
to $z=1090$, which depends strongly on the spatial curvature via
the hyperbolic sine in \re{bidt} \cite{Clarkson:2011d}. (Note
that the labels for the two curvature parameters
in table 1 of \cite{Clarkson:2011d} should be swapped.)
However, if the universe is not well described by an FLRW model,
it is possible that spatial curvature evolves so
that it is only significant at late times, and is not strongly
constrained by high-redshift probes \cite{Boehm:2013}.
Second, the large-angle anisotropy of the CMB is sensitive to
spatial curvature via the late Integrated Sachs-Wolfe (ISW) effect,
which is particularly important in \cite{Okouma:2012}.
However, analysis of the ISW effect depends on assumptions about
evolution of dark energy perturbations, which are rather speculative,
particularly if the equation of state crosses $-1$.

Model-independent constraints based only on geometrical optics, such
as the ones provided here or obtained from comparison of distance
with expansion rate
\cite{Shafieloo:2009, Mortsell:2011, Sapone:2014, Heavens:2014}
or cosmic parallax \cite{Rasanen:2013}, are thus complementary
to model-specific analyses, which involve more assumptions about the
matter content and the theory of gravity.

\para{Future constraints.}

In addition to strong lensing image deformation
by galaxies, existing observations of time delays and
both strong and weak lensing by galaxy groups and clusters
can be used to improve the constraints. Strong lensing by clusters
may be promising, because individual lenses have several
sources and some of the lenses are tightly modelled.
On the other hand, many source redshifts
are higher than current independent measurements of $d_s$.
The Euclid satellite, set to launch in 2020,
is expected to observe $10^5$ strong
lensing systems \cite{Euclid}. The usefulness of these systems
for the test discussed here depends on follow-up observations
to determine lens properties.
Better understanding of the systematics of modelling lensing
systems will be crucial.
Given such progress, we can expect constraints on deviations from
the FLRW metric, and on the spatial curvature of the FLRW universe,
to significantly improve in the near future from the proof
of concept we have presented here.
Assuming lens model Ia and a spatially flat FLRW model with
dust and vacuum energy, $10^4$ SNe \cite{LSST} and $10^4$
lensing data points with current errors give the constraint
$-0.03<\OKn<0.04$, within a factor of a few of the
current model-dependent range.

\para{Acknowledgments.}

We thank Artem Kupri for help with selecting and analysing
the lensing systems, Josiah Schwab for providing the measured
values of $\d$ and Adam Bolton for correspondence.
AF acknowledges support from the Finnish Academy, Grant No. 266918.
KB thanks the Australian Research Council for support through
the Future Fellowship (Grant No. FT140101270).
Computational resources used in this work were provided by Intersect
Australia Ltd.

\newpage

\begin{table}
\begin{tabular}{|c|cccccccccc|}
\hline
Name & $z_\l$ & $z_\s$ & $\sigma$ [km/s] & $\Delta\sigma$ [km/s]  & $\tE$ [as] & $\sigma_{\mathrm{atm}}$ & $\d$(3.0 as)  & $\d$(3.6 as) & $\d$(4.2 as) & Outlier in model Ia \\
\hline
J0029--0055 & 0.2270 & 0.9313 & 229 & 18 & 0.92 & 1.84000 & 2.355 & 2.391 & 2.422 & No \\
J0252+0039  & 0.2803 & 0.9818 & 164 & 12 & 1.04 & 2.07250 & 2.426 & 2.547 & 2.644 & Yes \\
J0405--0455 & 0.0753 & 0.8098 & 160 & 8 &  0.80 & 1.80527 & 2.484 & 2.502 & 2.522 & No \\
J0728+3835  & 0.2058 & 0.6877 & 214 & 11 & 1.25 & 1.80527 & 2.397 & 2.442 & 2.485 & Yes \\
J0737+3216  & 0.3223 & 0.5812 & 338 & 17 & 1.00 & 2.32818 & 2.278 & 2.328 & 2.373 & Yes \\
J0822+2652  & 0.2414 & 0.5941 & 259 & 15 & 1.17 & 1.93500 & 2.391 & 2.432 & 2.462 & No \\
J0912+0029  & 0.1642 & 0.3239 & 326 & 16 & 1.63 & 2.80200 & 2.162 & 2.228 & 2.289 & No \\
J0936+0913  & 0.1897 & 0.5880 & 243 & 12 & 1.09 & 1.43750 & 2.319 & 2.349 & 2.392 & No \\
J0946+1006  & 0.2219 & 0.6085 & 263 & 21 & 1.38 & 1.23667 & 2.321 & 2.386 & 2.437 & No \\
J0956+5100  & 0.2405 & 0.4699 & 334 & 17 & 1.33 & 1.65333 & 2.318 & 2.380 & 2.445 & No \\
J1023+4230  & 0.1912 & 0.6960 & 242 & 15 & 1.41 & 1.85000 & 2.367 & 2.428 & 2.477 & No \\
J1106+5228  & 0.0955 & 0.4069 & 262 & 13 & 1.23 & 1.94000 & 2.435 & 2.466 & 2.495 & Yes \\
J1153+4612  & 0.1797 & 0.8751 & 226 & 15 & 1.05 & 1.60667 & 2.502 & 2.545 & 2.580 & No \\
J1204+0358  & 0.1644 & 0.6307 & 267 & 17 & 1.31 & 1.41000 & 2.410 & 2.440 & 2.474 & No \\
J1205+4910  & 0.2150 & 0.4808 & 281 & 14 & 1.22 & 2.26800 & 2.316 & 2.345 & 2.381 & No \\
J1250+0523  & 0.2318 & 0.7953 & 252 & 14 & 1.13 & 2.10333 & 2.309 & 2.384 & 2.438 & No \\
J1402+6321  & 0.2046 & 0.4814 & 267 & 17 & 1.35 & 2.37000 & 2.319 & 2.358 & 2.393 & No \\
J1403+0006  & 0.1888 & 0.4730 & 213 & 17 & 0.83 & 1.77000 & 2.399 & 2.440 & 2.496 & No \\
J1420+6019  & 0.0629 & 0.5351 & 205 & 10 & 1.04 & 2.08333 & 2.391 & 2.443 & 2.487 & No \\
J1430+4105  & 0.2850 & 0.5753 & 322 & 32 & 1.52 & 1.58333 & 2.229 & 2.288 & 2.360 & No \\
J1531--0105 & 0.1596 & 0.7439 & 279  &14 & 1.71 & 1.83250 & 2.342 & 2.378 & 2.426 & No \\
J1538+5817  & 0.1428 & 0.5312 & 189 & 12 & 1.00 & 1.46000 & 2.386 & 2.448 & 2.507 & Yes \\
J1627--0053 & 0.2076 & 0.5241 & 290 & 15 & 1.23 & 1.85000 & 2.367 & 2.411 & 2.451 & No \\
J1630+4520  & 0.2479 & 0.7933 & 276 & 16 & 1.78 & 1.54333 & 2.360 & 2.416 & 2.469 & No \\
J1636+4707  & 0.2282 & 0.6745 & 231 & 15 & 1.09 & 1.35333 & 2.446 & 2.476 & 2.505 & No \\
J2238--0754 & 0.1371 & 0.7126 & 198 & 11 & 1.27 & 1.80527 & 2.324 & 2.374 & 2.418 & Yes \\
J2300+0022  & 0.2285 & 0.4635 & 279 & 17 & 1.24 & 1.93333 & 2.390 & 2.444 & 2.489 & No \\
J2303+1422  & 0.1553 & 0.5170 & 255 & 16 & 1.62 & 1.48333 & 2.214 & 2.272 & 2.330 & No \\
J2321--0939 & 0.0819 & 0.5324 & 249 & 12 & 1.60 & 1.58000 & 2.191 & 2.229 & 2.264 & No \\
J2341+0000  & 0.1860 & 0.8070 & 207 & 13 & 1.44 & 1.41667 & 2.140 & 2.183 & 2.239 & Yes \\
\hline
\end{tabular}

\caption{Data for the lensing systems. The quantity
$\Delta\sigma$ is the 1$\sigma$ error of the velocity
dispersion $\sigma$, $\sigma_{\mathrm{atm}}$ is the seeing
value, $\d$(n as) is the value of $\delta$ estimated
for an aperture of $n$ as. Values are from [19],
except for the seeing, which is from [20] and for
the values of delta, which are from [21].
For the three systems J0405-0455, J0728+3835 and J2238-0754,
no seeing value is given in [20] so we use the
average of the seeing values of the other 27 systems.
}

\label{tab:data}

\end{table}


\begin{thebibliography}{99}

\footnotesize{

\bibitem{Ferrer} F.~Ferrer and S.~R\"{a}s\"{a}nen,
\newblock JHEP02(2006)016
\newblock [arXiv:hep-th/0509225]
\nonum F.~Ferrer, T.~Multam\"{a}ki and S.~R\"{a}s\"{a}nen,
\newblock JHEP04(2009)006
\newblock [arXiv:0812.4182 [hep-th]]
\nonum F.~Ferrer,
\newblock {\it Nucl. Phys. Proc. Suppl.}  {\bf 194} (2009) 218
\newblock [arXiv:0907.1342 [hep-th]]

\bibitem{LTB} K. Enqvist,
\newblock \GRG{40} (2008) 451
\newblock \brt{arXiv:0709.2044} [astro-ph]
\nonum S.~February, J.~Larena, M.~Smith and C.~Clarkson,
\newblock \MNRAS{405} (2010) 2231
\newblock [arXiv:0909.1479 [astro-ph.CO]]
\nonum K.~Bolejko, M.-N.~ C\'el\'erier, A.~Krasinski,
\newblock \CQG{28} (2011) 164002
\newblock [arXiv:1102.1449]
\nonum M.~Redlich, K.~Bolejko, S.~ Meyer, G.F.~ Lewis and M.~Bartelmann,
\newblock \AaA{570} (2014) A63
\newblock [arXiv:1408.1872 [astro-ph.CO]]

\bibitem{BR} S. R\"{a}s\"{a}nen,
\newblock JCAP02(2009)011
\newblock \brt{arXiv:0812.2872 [astro-ph]}
\nonum S. R\"{a}s\"{a}nen,
\newblock JCAP03(2010)018
\newblock [arXiv:0912.3370 [astro-ph.CO]]
\nonum T.~Buchert and S.~R\"{a}s\"{a}nen,
\newblock {\it Ann. Rev. Nucl. Part. Sci.} {\bf 62} (2012) 57q
\newblock [arXiv:1112.5335 [astro-ph.CO]]
\nonum M.~Lavinto, S.~R\"{a}s\"{a}nen and S.J.~Szybka,
\newblock JCAP12(2013)051
\newblock [arXiv:1308.6731 [astro-ph.CO]]

\bibitem{Boehm:2013} C.~Boehm and S. R\"{a}s\"{a}nen,
\newblock JCAP09(2013)003 
\newblock [arXiv:1305.7139 [astro-ph.CO]]

\bibitem{Clarkson:2007b} C. Clarkson, B.A. Bassett and T.C. Lu,
\newblock \PRL{101} (2008) 011301
\newblock \brt{arXiv:0712.3457 [astro-ph]}

\bibitem{Shafieloo:2009} A.~Shafieloo and C.~Clarkson,
\newblock \PRD{81} (2010) 083537
\newblock [arXiv:0911.4858 [astro-ph.CO]]

\bibitem{Mortsell:2011} E.~M\"ortsell and J.~J\"onsson,
\newblock [arXiv:1102.4485 [astro-ph.CO]]

\bibitem{Sapone:2014} D.~Sapone, E.~Majerotto and S.~Nesseris,
\newblock \PRD{90} (2014) 023012
\newblock [arXiv:1402.2236 [astro-ph.CO]]

\bibitem{Rasanen:2013} S.~R\"{a}s\"{a}nen,
\newblock JCAP03(2014)035
\newblock [arXiv:1312.5738 [astro-ph.CO]]

\bibitem{Peebles:1993} P.J.E. Peebles,
\newblock {\it Principles of Physical Cosmology},
\newblock 1993, Princeton University Press

\bibitem{Bernstein:2005} G.~Bernstein,
\newblock \APJ{637} (2006) 598
\newblock [arXiv:astro-ph/0503276]

\bibitem{Suzuki:2011} N.~Suzuki et al. (The Supernova Cosmology Project),
\newblock \APJ{746} (2012) 85
\newblock [arXiv:1105.3470 [astro-ph.CO]]

\bibitem{duality} I.M.H. Etherington,
\newblock {\it Philosophical Magazine} {\bf 15} (1933) 761
\newblock Reprinted in \GRG{39} (2007) 1055
\nonum G.F.R. Ellis,
\newblock {\it Relativistic Cosmology}, 1971,
\newblock General Relativity and Cosmology, ed. R.K. Sachs,
\newblock Academic Press Inc., London,
\newblock p. 104,
\newblock Reprinted in \GRG{41} (2009) 581

\bibitem{Nadathur:2010} S.~Nadathur and S.~Sarkar,
\newblock \PRD{83} (2011) 063506
\newblock [arXiv:1012.3460 [astro-ph.CO]]

\bibitem{fitter} G.R.~Bengochea,
\newblock \PLB{696} (2011) 5
\newblock [arXiv:1010.4014 [astro-ph.CO]]
\nonum Z.~Li, P.~Wu and H.~Yu,
\newblock JCAP11(2010)031
\newblock [arXiv:1011.2036 [gr-qc]]
\nonum M.C.~March, R.~Trotta, P.~Berkes, G.D.~Starkman and P.M.~Vaudrevange,
\newblock \MNRAS{418} (2011) 2308
\newblock  [arXiv:1102.3237 [astro-ph.CO]]
\nonum R.~Giostri, M. Vargas dos Santos, I.~Waga, R.R.R.~Reis, M.O.~Calvao and B.L.~Lago,
\newblock JCAP03 (2012) 027
\newblock [arXiv:1203.3213 [astro-ph.CO]]

\bibitem{Kochanek:2004} C.S. Kochanek, P. Schneider and J. Wambsganss,
\newblock {\it The Saas Fee Lectures on strong gravitational lensing},
\newblock 2004, published in part 2 of ``Gravitational Lensing: Strong, Weak \& Micro'', proceedings of the 33rd Saas-Fee Advanced Course, G. Meylan, P. Jetzer and P. North (eds.), Springer-Verlag: Berlin
\newblock [arXiv:astro-ph/0407232]

\bibitem{Kochanek:1999} C.S.~Kochanek \etal,
\newblock \APJ{543} (2000) 131
\newblock [arXiv:astro-ph/9909018]

\bibitem{Ofek:2003} E.O.~Ofek, H.-W.~Rix and D.~Maoz,
\newblock \MNRAS{343} (2003) 639
\newblock [arXiv:astro-ph/0305201]

\bibitem{Schwab:2009} J.~Schwab, A.S.~Bolton and S.A.~Rappaport,
\newblock \APJ{708} (2010) 750
\newblock [arXiv:0907.4992 [astro-ph.CO]]

\bibitem{table2} See table \ref{tab:data}.

\bibitem{Bolton:2008} A.S.~Bolton \etal,
\newblock \APJ{682} (2008) 964
\newblock [arXiv:0805.1931 [astro-ph]]

\bibitem{SIMBAD} http://simbad.u-strasbg.fr/simbad/sim-fid

\bibitem{Schwab:2014} J. Schwab, private communication.

\bibitem{recon} C. Catto\"{e}n and M. Visser,
\newblock \brt{arXiv:gr-qc/0703122}
\nonum S.~Nesseris and J.~Garc\'{i}a-Bellido,
\newblock \PRD{88} (2013)  063521
\newblock [arXiv:1306.4885 [astro-ph.CO]]

\bibitem{CMB} M.~Vonlanthen, S.~R\"{a}s\"{a}nen and R.~Durrer,
\newblock JCAP08(2010)023
\newblock [arXiv:1003.0810 [astro-ph.CO]]
\nonum B.~Audren, J.~Lesgourgues, K.~Benabed and S.~Prunet,
\newblock JCAP02(2013)001
\newblock [arXiv:1210.7183 [astro-ph.CO]]
\nonum B.~Audren,
\newblock \MNRAS{444} (2014) 827
\newblock [arXiv:1312.5696 [astro-ph.CO]]

\bibitem{Efstathiou:2013} G.~Efstathiou,
\newblock \MNRAS{440} (2014) 1138
\newblock [arXiv:1311.3461 [astro-ph.CO]]

\bibitem{Press:2007} W.H. Press, S.A. Teukolsky, W.T. Vetterling and B.P. Flannery,
\newblock Numerical Recipes 3rd Edition: The Art of Scientific Computing,
\newblock 2007,
\newblock Cambridge University Press, New York

\bibitem{Planck:cosmo} P.A.R.~Ade et al. [Planck Collaboration],
\newblock \AaA{571} (2014) A16
\newblock [arXiv:1303.5076 [astro-ph.CO]]

\bibitem{Okouma:2012} P.M.~Okouma, Y.~Fantaye and B.A.~Bassett,
\newblock \PLB{719} (2013) 1
\newblock [arXiv:1207.3000 [astro-ph.CO]]

\bibitem{Busti:2014} V.C.~Busti, C.~Clarkson and M.~Seikel,
\newblock \MNRAS{441} (2014) L11
\newblock [arXiv:1402.5429 [astro-ph.CO]]

\bibitem{Heavens:2014} A.~Heavens, R.~Jimenez and L.~Verde,
\newblock \PRL{113} (2014) 241302
\newblock [arXiv:1409.6217 [astro-ph.CO]

\bibitem{Clarkson:2011d} C.~Clarkson, T.~Clifton, A.~Coley and R.~Sung,
\newblock \PRD{85} (2012) 043506
\newblock [arXiv:1111.2214 [astro-ph.CO]]

\bibitem{Euclid} http://sci.esa.int/euclid/

\bibitem{LSST} http://lsst.org/lsst/science/scientist\_supernovae/

}

\end{thebibliography}
\end{document}